\documentstyle[11pt,epsfig]{article}

\newcommand{\BABARPubYear}    {01}

\newcommand{\BABARProcNumber} {07}
\newcommand{\SLACPubNumber} {8802}

\newcommand{\vs}{\\[0.3cm]\noindent}

\input pubboard/babarsym

\def\Btopsikst     {{ B \to \jpsi \Kstar}}
\def\Btopsikstz    {{\Bz \to \jpsi \Kstarz}}
\def\Btopsikstp    {{\Bu \to \jpsi \Kstarp}}

\def\ksttokspiz {{\Kstarz \to \KS \piz}}
\def\ksttokpi   {{\Kstarz \to \Kpm \pimp}}

\def\az    {{A_{0}}}
\def\ap    {{A_{\parallel}}}
\def\at    {{A_{\perp}}}
\def\azd   {{|\az|^{2}}}
\def\apd   {{|\ap|^{2}}}
\def\atd   {{|\at|^{2}}}

\def\phip  {{\phi_{\parallel}}}
\def\phit  {{\phi_{\perp}}}
\def\thetakstar {{\theta_{\Kstar}}}
\def\phitr      {{\phi_{\rm tr}}}
\def\thetatr    {{\theta_{\rm tr}}}
\def\cthetatr   {{\cos{\thetatr}}}

\def\cthetakstar{{\cos{\thetakstar}}}

\def\sphitr     {{\sin{\phitr}}}
\def\pipt       {{\rm{Im}{(\ap^{*}\at)}}}
\def\przp       {{\rm{Re}{(\az^{*}\ap)}}}
\def\pizt       {{\rm{Im}{(\az^{*}\at)}}}

\newcommand{\cq}[1]{\cos^{2}{#1}}
\newcommand{\sq}[1]{\sin^{2}{#1}}
\newcommand{\gfrac}[2]{\displaystyle\frac{#1}{#2}}
\newcommand{\dd}{\rm{d}}

\def\gobs       {{g_{\rm obs}}}

\def\Journal#1#2#3#4{{#1} {\bf #2}, #3 (#4)}

\def\NIMA{{\em Nucl. Instrum. Methods} A}
\def\NPB{{\em Nucl. Phys.} B}
\def\PLB{{\em Phys. Lett.}  B}
\def\PRL{\em Phys. Rev. Lett.}
\def\PRD{{\em Phys. Rev.} D}
\def\ZPC{{\em Z. Phys.} C}

\def\ra{\rightarrow}

\def\be{\begin{equation}}
\def\ee{\end{equation}}
\def\bea{\begin{eqnarray}}
\def\eea{\end{eqnarray}}

\long\def\inst#1{\par\nobreak\kern 4pt\nobreak
    {\it #1}\par\vskip 10pt plus 3pt minus 3pt}

\begin{document}
{\pagestyle{empty}

\begin{flushright}
SLAC-PUB-\SLACPubNumber \\
\babar-PROC-\BABARPubYear/\BABARProcNumber \\
March, 2001 \\
\end{flushright}

\par\vskip 3cm

\begin{center}
\Large \bf Hadronic \boldmath$B$ Decays to Charmless Final States
	and to \boldmath$J/\psi K^*$
\end{center}
\bigskip

\begin{center}
\large 
A.~H\"ocker \\
Laboratoire de l'Acc\'el\'erateur Lin\'eaire,\\
	IN2P3-CNRS et Universit\'e de Paris-Sud, \\
	BP 34, F-91898 Orsay Cedex, France\\
	E-mail: hoecker@lal.in2p3.fr\\
(for the \lbabar\ Collaboration)
\end{center}
\bigskip \bigskip

\begin{center}
\large \bf Abstract
\end{center}
Preliminary results 
from the \babar\ experiment on charmless $\B$ decays
to charged pions or kaons,
and the measurement of the $\Btopsikst$ decay amplitudes
are presented. The data sample, collected at the 
asymmetric-energy $\B$-factory PEP-II at SLAC,
comprises a total number of
22.7 million $\FourS$ decays, corresponding to an
integrated on-resonance luminosity of approximately 
$21~{\rm fb}^{-1}$. We measure the following $\CP$-averaged
branching fractions:
$\BR(\Bztopipi)
=(4.1\pm1.0({\rm stat})\pm0.7({\rm sys}))\times10^{-6}$,
$\BR(\Bztokpi)
=(16.7\pm1.6({\rm stat})^{\,+1.2}_{\,-1.7}({\rm sys}))\times10^{-6}$,
and an upper limit of $\BR(\Bztokk)<2.5\times10^{-6}$,
at 90\% confidence limit.
The measurement of the $J/\psi K^*$ decay amplitudes results in
$R_\perp=0.160\pm0.032({\rm stat})\pm0.036({\rm sys})$,
and reveals a dominant longitudinal component. The phase
of the longitudinal amplitude shows evidence for
non-vanishing final-state interaction.

\vfill
\begin{center}
Contributed to the Proceedings of the \\
4th International Conference on B Physics and $\CP$
Violation - BCP4 \\
2/13/2000---2/19/2000, Ise-Shima, Japan
\end{center}

\vspace{1.0cm}
\begin{center}
{\em Stanford Linear Accelerator Center, Stanford University, 
Stanford, CA 94309} \\ \vspace{0.1cm}\hrule\vspace{0.1cm}
Work supported in part by Department of Energy contract DE-AC03-76SF00515.
\end{center}


%
%
\section{Charmless Hadronic $\B$ Decays}
\label{sec:charmlessHadronic}

Hadronic $\B$ decays to charmless final states provide
a rich environment for $\CP$ violation studies as well as 
searches for new physics. Any $b\rightarrow u\bar{u}d$
transitions, where the hadronic final state may be a $\CP$
eigenstate or not, represents a possible source for measuring
the angle $\alpha={\rm arg}[-V_{td}V_{tb}^*/V_{ud}V_{ub}^*]$ 
of the unitarity triangle (UT). Indirect
$\CP$ violation in $\BzBzb$ mixing results from the 
interference between direct and mixed decays. The weak 
phase $\alpha$ can be determined from the measurement of
the time-dependent asymmetry of $\Bztopipi$, together with 
constraints from isospin conservation~\cite{ref:btopipi}.
Moreover, a three-body Dalitz plot analysis of $B \to \rho\pi$ 
decays allows one to simultaneously fit for $\alpha$ and the
contributing strong tree and penguin phases~\cite{ref:snyder}.
Direct $\CP$ violation arises from the interference between 
tree and penguin amplitudes contributing to the hadronic
final state. It can be measured in charged as well as
neutral $\B$ decays. Several charmless modes 
are self-tagging which facilitates the search for direct
$\CP$ violation. Theories employing perturbative 
QCD~\cite{ref:sanda} or non-leading Factorization 
Approximation (FA)~\cite{ref:beneke} have been shown to provide
predictions of charmless branching fractions of tree and
penguins dominated $\B$ decay modes. Their measurement
can exhibit tests of the theory and may eventually be used
to determine the UT angle 
$\gamma={\rm arg}(V_{ub}^*)$~\cite{ref:fleischer}.
Finally, the $\Bztokk$ branching fraction
provides a measure of contributions from $W$ exchange 
and hard rescattering topologies which are suppressed in the FA.
\vs
Preliminary $\babar$ results on charmless $\B$ decays using 
approximately half of the present data set have been 
presented in Ref.~\cite{ref:theresa}. Significant signals
have been found for the modes: $\Bztopipi$, 
$\Bztokpi$, $\Bztorhopi$ and $B^0 \to \eta^\prime K$.
Updates of these analyses as well as measurements of
a large number of new charmless modes are being prepared
within the $\babar$ Collaboration. Important activity 
is also invested into the analyses of 
$\btosgam$ decays for which results are forthcoming.
\vs
In this note we will concentrate on the measurement of
the charged two-body modes $B^0 \to h^+h^{\prime-}$, 
with $h^{(\prime)}$ being pion or kaon.  We use (almost)
the full year 1999-2000 data set which amounts to 
$(22.57\pm0.36)\times10^{6}$  $B\Bbar$ pairs, corresponding
to an integrated luminosity of approximately 
$21~{\rm fb}^{-1}$ collected by the $\babar$ 
detector~\cite{ref:babardet} at the PEP-II storage 
ring. Background studies utilize approximately
$3~{\rm fb}^{-1}$ of data taken below the $B\Bbar$ threshold.
A ``blind'' analysis methodology has been adopted throughout,
so that the signal region for each mode remained hidden
until finalization of event selection and systematic 
studies.

\subsection{$B^0 \to h^+h^{\prime-}$ Candidate Selection}

We impose efficient track quality requirements on charged
tracks which must have a minimum transverse momentum of 
$100\mevc$ in the laboratory frame (LAB). Charged pions and 
kaons are identified by measuring the \v{C}erenkov
angle, $\theta_C$, of photons produced while traversing 
a radiative medium made of synthetic fused silica, 
the DIRC~\cite{ref:dirc}, surrounding the drift 
chamber in the barrel~\cite{ref:babardet}. 
\vs
We select $B^0 \to h^+h^{\prime-}$ candidates based on the 
energy-substituted mass,
$\mes$, where $\sqrt{s}/2$ ($\sqrt{s}$ being the center-of-mass
energy) is substituted for the candidate's 
energy, and the difference $\Delta E$ between the $\B$-candidate 
energy and $\sqrt{s}/2$. Preselected $\B$ candidates are
required to satisfy $5.2<\mes<5.3~\gevcc$ and 
$|\Delta E|<0.42~\gev$. The dominant background for all charmless 
modes stems from continuum $q\bar{q}$ production, which 
exhibits a jet-like structure that distinguishes it from 
the more spherical topology of generic $B\Bbar$ events. 
To suppress this background we use a Fisher discriminant,
${\cal F}$, combining the charged and neutral momentum flow from the 
rest of the event relative to the candidate's thrust axis, 
characterized by nine concentric, $10^\circ$ wide 
cones~\cite{ref:cleofisher}.
\begin{figure}
\center
\hspace{-0.4cm}
\epsfig{file=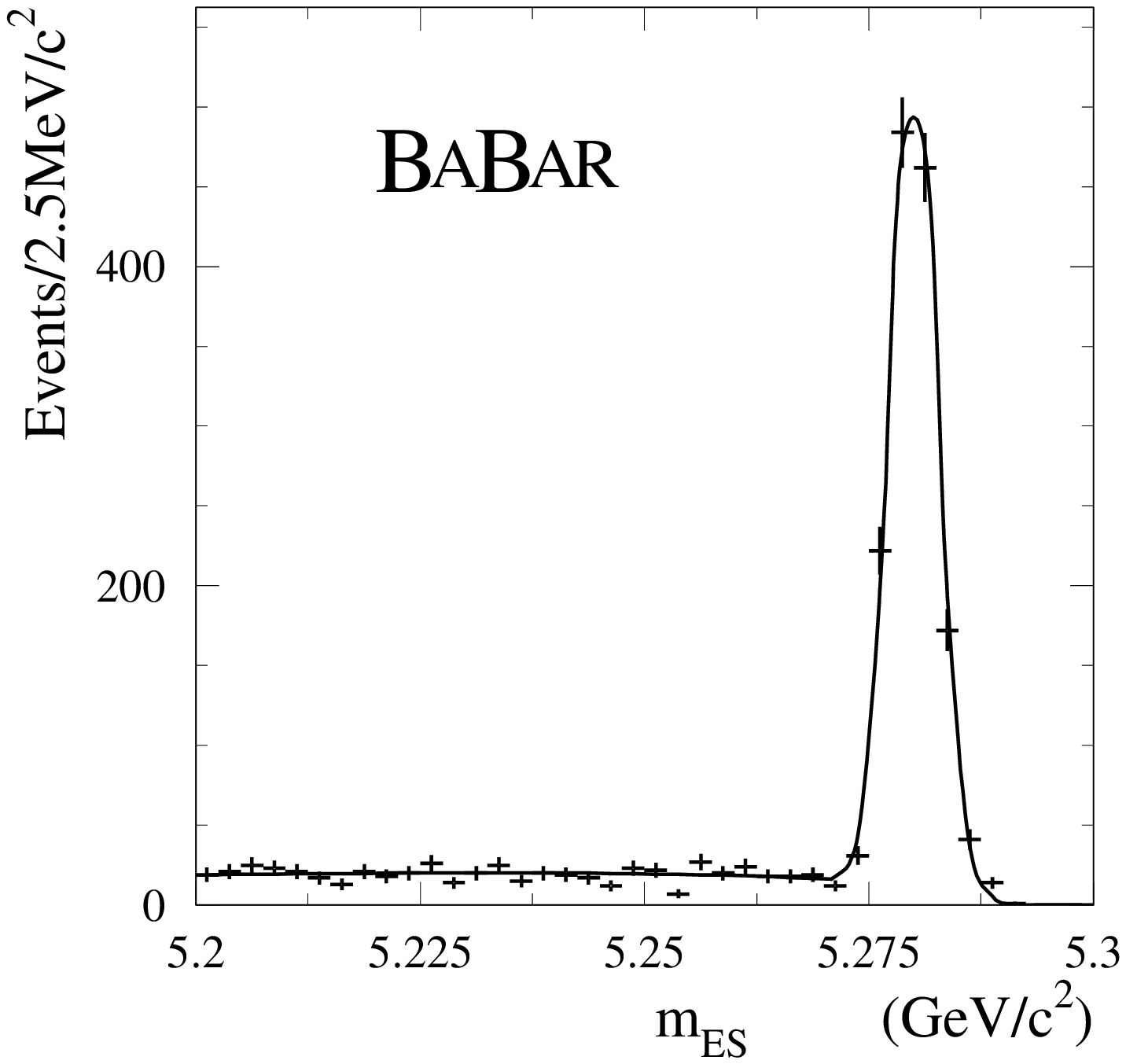,width=3in}
\hspace{-0.0cm}
\epsfig{file=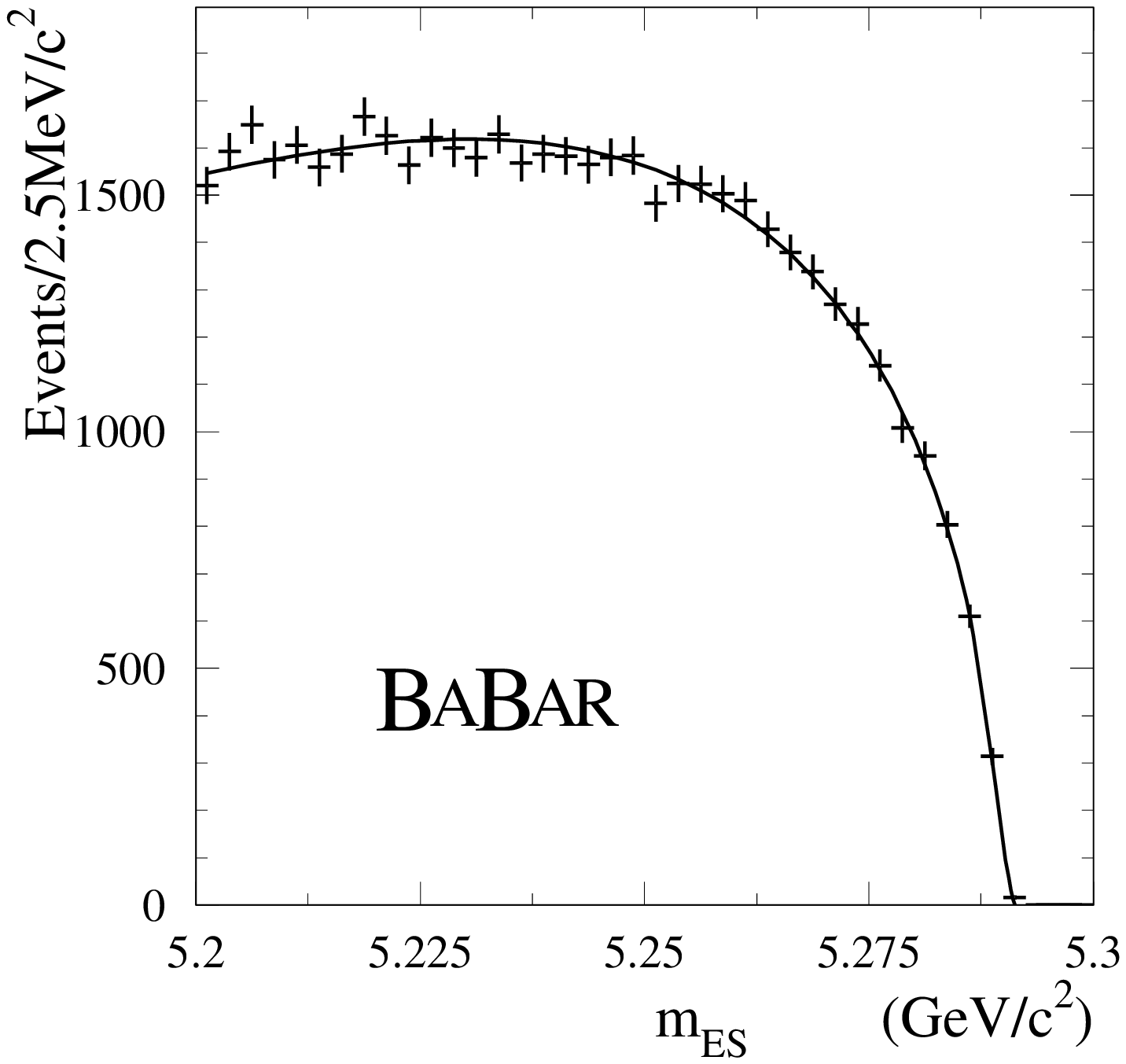,width=3in}
\caption[.]{Energy substituted mass spectra. Left: 
	$B^- \to D^0\pi^-$ signal with single Gaussian fit.
	Right: background from $\Delta E$ sidebands
	of $B\to h^+ h^{\prime-}$ candidates, fit to an ARGUS
	shape function~\cite{ref:argusbkg}.}
\label{fig:brmescalib}
\end{figure}

\subsection{Calibration of the Discriminating Variables}

The signal and background distributions of the 
discriminating variables are controlled by independent
calibration samples obtained from: $\Delta E$ and
$\mes$ sidebands and off-resonance data to study
background characteristics, 
exclusive $B^- \to D^0\pi^-\to(K^-\pi^+)\pi^+$
decays to gauge the $\Delta E$ and $\mes$ resolution
of the Monte Carlo (MC) simulation with respect to the
data, inclusive $D^{*+}\to D^0\pi^+\to (K^-\pi^+)\pi^+$
decays to measure the particle identification (PID)
performance of the DIRC, and finally MC simulated
signal and background events. The calibration measures
a $\mes$ resolution of $2.6~\mevcc$ for signal events 
(see Fig.~\ref{fig:brmescalib} for signal and background
distributions), and a $\Delta E$ resolution of 
approximately $26~\mev$. The latter
value has been rescaled to account for differences 
between data and MC simulation with a conservative error 
estimated to be $\pm5~\mev$. The output 
of the Fisher discriminant is found to be well reproduced
by the MC simulation. A sample of $D^*$-tagged
$D^0\to K^-\pi^+$ decays is used to parameterize 
the $\theta_C$ distributions for pion and kaon tracks 
as a function of the track's polar angle. The $K/\pi$ separation
varies from $2$ to more than $8$ standard deviations 
across the relevant momentum range between $1.7$
and $4.2~\gevc$ in the LAB.  

\subsection{Likelihood Fit}

Signal yields in all three $h^+h^{\prime-}$ modes are determined
simultaneously from an 8 parameter, unbinned maximum likelihood (UML) 
fit, incorporating $\mes$, $\Delta E$, ${\cal F}$, and the measured 
$\theta_C$ for each track. All candidates passing the
preselection are included in the fit. The probability density
function (PDF) for an event $i$, with a number of signal 
(background) candidates, $N_{({\rm b})hh^\prime}$, is given by
\begin{equation}
	{\cal P}_i(\mu)
	\propto\! \sum_{h\le h^\prime=\pi,K}\!
		\left(
		N_{hh^\prime}{\cal P}_i^{hh^\prime}(\mu)
	  +\, N_{{\rm b}hh^\prime}{\cal P}_i^{{\rm b}hh^\prime}(\mu)
		\right) 
	\;+\;{\rm asymmetries}
\end{equation}
with $(\mu)=(\mes, \Delta E, {\cal F}, \theta_{C,,}^+, \theta_{C,,}^-)$
being the discriminating variables that enter the UML fit.
The combined PDF, ${\cal P}^{{\rm (b)}hh^\prime}(\mu)$, is the 
product of the PDF's of the individual variables. The total 
extended log-likelihood~\cite{ref:barlow}
for $N$ candidates, which is to be maximized in the fit, reads
\begin{equation}
	{\cal L} = \sum_{i=1}^{N}{\cal P}_i(\mu) -N^\prime,
\end{equation}
where $N^\prime=\sum_{h,h^\prime}(N_{h,h^\prime}+N_{{\rm b}hh^\prime})$
is the Poissonian extension of the likelihood.
A bias, enhancing the contributions from $\pi^+\pi^-$ 
and $K^+K^-$ with respect to $K^+\pi^-$, has been 
identified in the 6 parameter UML fit previously 
applied~\cite{ref:theresa}. This was due to the implicit 
assumption of randomly paired background tracks. The addition
of 2 independent background parameters provides the remedy to 
the problem. 

\subsection{Systematic Uncertainties}

Systematics uncertainties arise from the imperfect knowledge
of the PDF shape parameterizations. Associated errors
are quantified by varying the parameters within their statistical
errors and to cover disagreements between MC and data. 
Alternate parameterizations are used in addition. 
A large number of cross checks have been performed
to validate the fit procedure. Toy MC studies showed no bias
in the fit outputs. A cut-based analysis, performed in parallel,
agrees with the results from the UML fit.

\subsection{Results}

With signal efficiencies of $\epsilon_{\pi\pi}=0.448\pm0.008$,
$\epsilon_{K\pi}=0.447\pm0.008$ and 
$\epsilon_{KK}=0.429\pm0.008$, we find signal yields
$N_{\pi\pi}=41\pm10\pm7$, 
$N_{K\pi}=169\pm17^{\,+12}_{\,-17}$ and
$N_{KK}=8.2^{\,+7.8}_{\,-6.4}\pm3.3$, from the likelihood
fit, where the uncertainties
are statistical and systematic.
The correlations between these numbers are below 15\%.
The systematic errors are dominated by the uncertainties in 
the description of ${\cal F}$ and the resolution on $\Delta E$.
Figure~\ref{fig:brmes} shows the $\mes$ 
distributions for candidates of the three final states from 
\begin{figure}[t]
\center
\epsfig{file=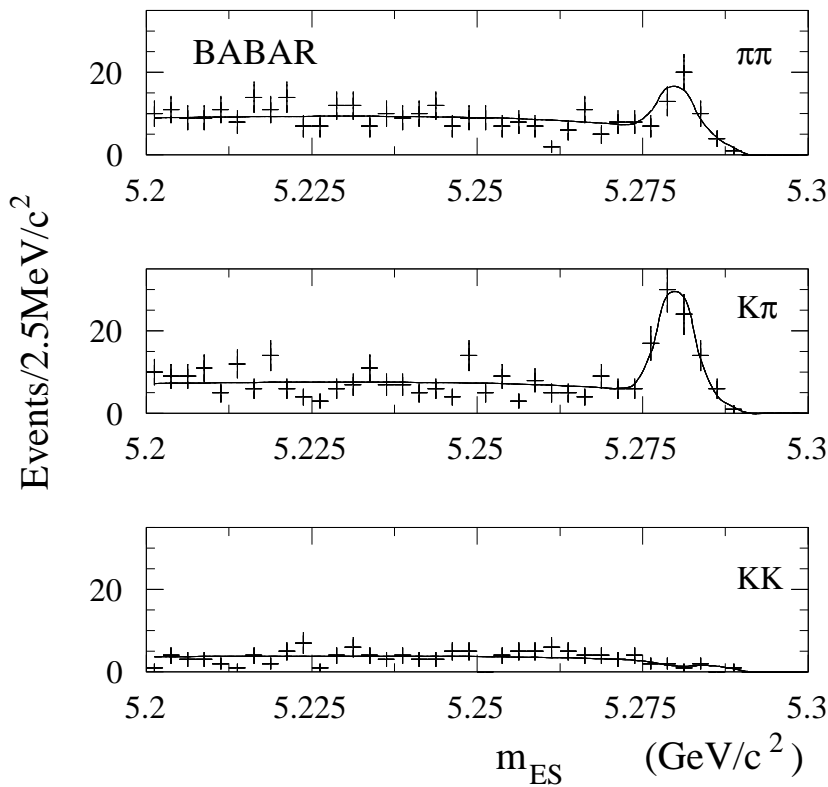,width=4.5in}
\vspace{-0.3cm}
\caption{Distribution of $\mes$ for the $B^0$ final states
	$\pi^+\pi^-$ (upper plot), 
	$K^+\pi^-$ (middle plot) and 
	$K^+K^-$ (lower plot), 
	from the cut based analysis.}
\label{fig:brmes}
\end{figure}
the cut based analysis.
\vs
The branching fractions are obtained from the signal yield {\em via} 
$\BR(B\to h^+h^{\prime-})
=N_{hh^{\prime}}/(\epsilon_{hh^{\prime}} N_{B\Bbar})$, giving
\begin{eqnarray}
	\BR(\Bztopipi) &\,=\,& 
		(4.1\pm1.0\pm0.7)\times10^{-6}~, \\
	\BR(\Bztokpi) &\,=\,&
		(16.7\pm1.6^{\,+1.2}_{\,-1.7})
		\times10^{-6}~, \\
	\BR(\Bztokk) &\,<\,&
		2.5\times10^{-6}~\hspace{0.3cm}(90\%~{\rm CL})~,
\end{eqnarray}
where the uncertainties are statistical and systematic.
The statistical significance is 4.7 (15.8) standard deviations
for the $\pi^+\pi^-$ ($K^+\pi^-$) final state. We do not find
a significant signal yield in the mode $\Bztokk$ and hence
quote an upper limit at 90\% confidence level.
The results are consistent with our previous preliminary 
numbers~\cite{ref:theresa}, based on less than half
the present data set. 
A comparison with published branching ratios 
from CLEO~\cite{ref:brcleo} 
and preliminary results from BELLE~\cite{ref:brbelle} 
is given in Fig.~\ref{fig:brcomp}, showing consistency.

\begin{figure}[t]
\center
\epsfig{file=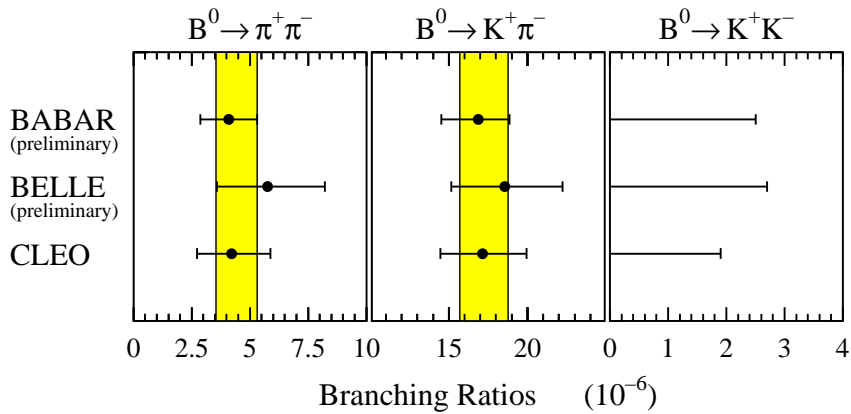,width=4.5in}
\caption{Comparison of measured 
	branching ratios to $h^+ h^{\prime-}$ final states.
	The shaded bands give the averages for the 
	significant measurements.}
\label{fig:brcomp}
\end{figure}
\hfill
\pagebreak
%
%
\section{Measurement of the $\Btopsikst$ Decay Amplitudes}

The decay $\Btopsikstz$, $\ksttokspiz$ can be used
in time dependent $\CP$ asymmetry studies
to measure the $\CP$ violating parameter $\stwob$.
In contrast to the pure $\CP$-odd eigenstate $\bpsiks$, 
this pseudoscalar to vector-vector decay
gives a mixture of $\CP$-odd and $\CP$-even eigenstates since 
orbital angular momenta $L = 0, 1, 2$
of the $(\jpsi , \Kstarz )$ system are allowed.
If one $\CP$ contribution dominates,
$\stwob$ can be measured as in the $\bpsiks$ case, using 
the time distribution only. Otherwise, 
both the angular and time distributions
are needed to avoid a partial cancellation of the observed
$\CP$ asymmetry~\cite{ref:dunietz}.
Assuming the weak decay to be isospin invariant,
the amplitudes can be measured
using all $\Kstar\to K\pi$ modes, which greatly increases
statistics.
\vs
Three amplitudes are necessary to describe the decay.
In the transversity basis~\cite{ref:dunietz,ref:dighe},
used in previous analyses by CLEO~\cite{ref:jkcleo} 
and CDF~\cite{ref:jkcdf}, the amplitudes  
$\ap$, $\az$ and $\at$, have definite $\CP$ eigenvalues: the 
latter is $\CP$-odd and the former two are $\CP$-even.
The longitudinal polarization fraction of the decay
is measured by $|\az|^2 \equiv \Gamma_L/\Gamma$, whereas
$\ap$ and $\at$ govern the transverse polarization and correspond 
to parallel and perpendicular polarizations of the two vector 
mesons, respectively. Hence, $|\at|^2$ measures the $P$ wave 
contribution to the decay.
The measurement of the decay amplitudes also provides a test 
of the factorization hypothesis. In the factorization
scheme, the weak decay is described by a product of two 
hadronic currents, $\jpsi$ and $\B \to \Kstar$, where
final-state interaction is neglected. If factorization 
holds, the corresponding amplitudes should have trivial 
relative phases of $0$ or $\pi$.
\vs
We present preliminary results on
a complete angular analysis in the transversity basis 
considering the decays $\Btopsikstz$ and $\Btopsikstp$, where 
the $\Kstarz$ and $\Kstarp$ are reconstructed in the 
modes $\KS\piz$, $\Kp\pim$ and $\KS\pip$ and $\Kp\piz$,
the $\KS$ decaying into two charged pions,
and the $\jpsi$ through decays into $\epem$ and $\mumu$. 
The data sample used corresponds to the full year 1999-2000 
recording of $(22.7\pm0.4)\times10^{6}$ $B\Bbar$ pairs
(see Section~\ref{sec:charmlessHadronic}).

\subsection{$\Btopsikst$ Candidate Selection}

Charged tracks are used in a fiducial polar angle range 
where PID efficiencies are well understood.
Electrons are identified requiring their momentum to be 
compatible with their energy deposition in the electromagnetic
calorimeter (EMC), and their $\dedx$ to be compatible with 
the electron hypothesis. Muons are
required to penetrate at least 2 interaction lengths in the 
detector and to leave a low number of hits per layer in the 
Instrumented Flux Return (IFR). 
If a muon penetrates the EMC, it is required to deposit energy 
consistent with a  minimum ionizing particle.
Kaon candidates are required to pass a pion veto based on 
DIRC and $\dedx$ information. 
\vs
$\jpsi$ candidates are selected requiring both leptons to 
be identified. The vertexed lepton pair is required to have 
an invariant mass in the range 3.06-3.14~$\gevcc$ for muons 
and 2.95-3.14~$\gevcc$ for electrons.
$\KS$ are reconstructed as vertexed pairs of charged 
tracks with an invariant mass between 489 and 507~$\mevcc$
and a flight direction which is compatible to 
originate from the interaction point.
Photons are defined as neutral clusters with an energy 
greater than 30~$\mev$ and photon-like shower shapes.
Neutral pions are selected from pairs of photons with an invariant 
mass between 106 and 153~$\mevcc$. The $\jpsi$, $\KS$ and 
$\piz$ are fitted to their respective nominal masses.
$\Kstar$ are formed as $K\pi$ combinations with an invariant 
mass within $\pm 100~\mevcc$ of the nominal $\Kstar(892)$ mass. 
\vs
Finally, $\B$ mesons are formed by combining $\jpsi$ 
and $\Kstar$ candidates. Furthermore, requirements 
on the cosine of the $\Kstar$ 
helicity angle reduce the cross feed (CF) from $\jpsi(K\pipm)^*$ 
modes, where the $\pipm$ is lost, and the self cross feed
(SCF) due to a wrongly reconstructed $\piz$ picked up by 
the reconstruction. The (S)CF is the most important 
background and accumulates in the $\mes$ signal region.
$\B$ meson candidates are selected with $\Delta E$
(required to be between $-70$~$\mev$ and $+50$~$\mev$ for channels
involving a $\piz$ and within $\pm 30$~$\mev$ in the other cases)
and $\mes$. 
For the $\mes$ signal region between 5.27 and 5.29~$\gevcc$ 
we obtain reconstruction efficiencies of 9.9\%, 23.9\%, 17.2\% 
and 13.8\% for the $\jpsi\KS\piz$, $\jpsi K^+\pi^-$, $\jpsi\KS\pi^+$ 
and $\jpsi K^+\piz$ modes, respectively.
The corresponding (S)CF contamination amounts to
(1.6)15.8\%, (0.6)2.4\%, (0.5)3.0\% and (2.2)15.7\% of the signal 
contributions.  The signal yields are obtained by a fit 
of the $\mes$ spectra describing the background by an ARGUS
function~\cite{ref:argusbkg} and the signal by a Gaussian:
$N\left(\jpsi\KS \piz \right)=39\pm 6$,
$N\left(\jpsi K^+\pi^-\right)=530\pm 32$,
$N\left(\jpsi\KS \pi^+\right)=131\pm 12$ and
$N\left(\jpsi K^+\piz \right)=195\pm 15$.
\vs
As shown in Ref.~\cite{ref:jkcleo}, $K\pi$ production other
than {\em via} the $\Kstar(892)$ contributes to $\B\to\jpsi K\pi$ 
final states. Figure~\ref{fig:heavykstar} shows the $K\pi$ mass 
distribution for the $\B\to\jpsi\Kpm\pimp$ mode, after subtraction 
of combinatoric $K\pi$ background from the low $\mes$ region.
We observe clear evidence of resonant excitation around 
1430~$\mevcc$.

\begin{figure}
\center
\epsfig{file=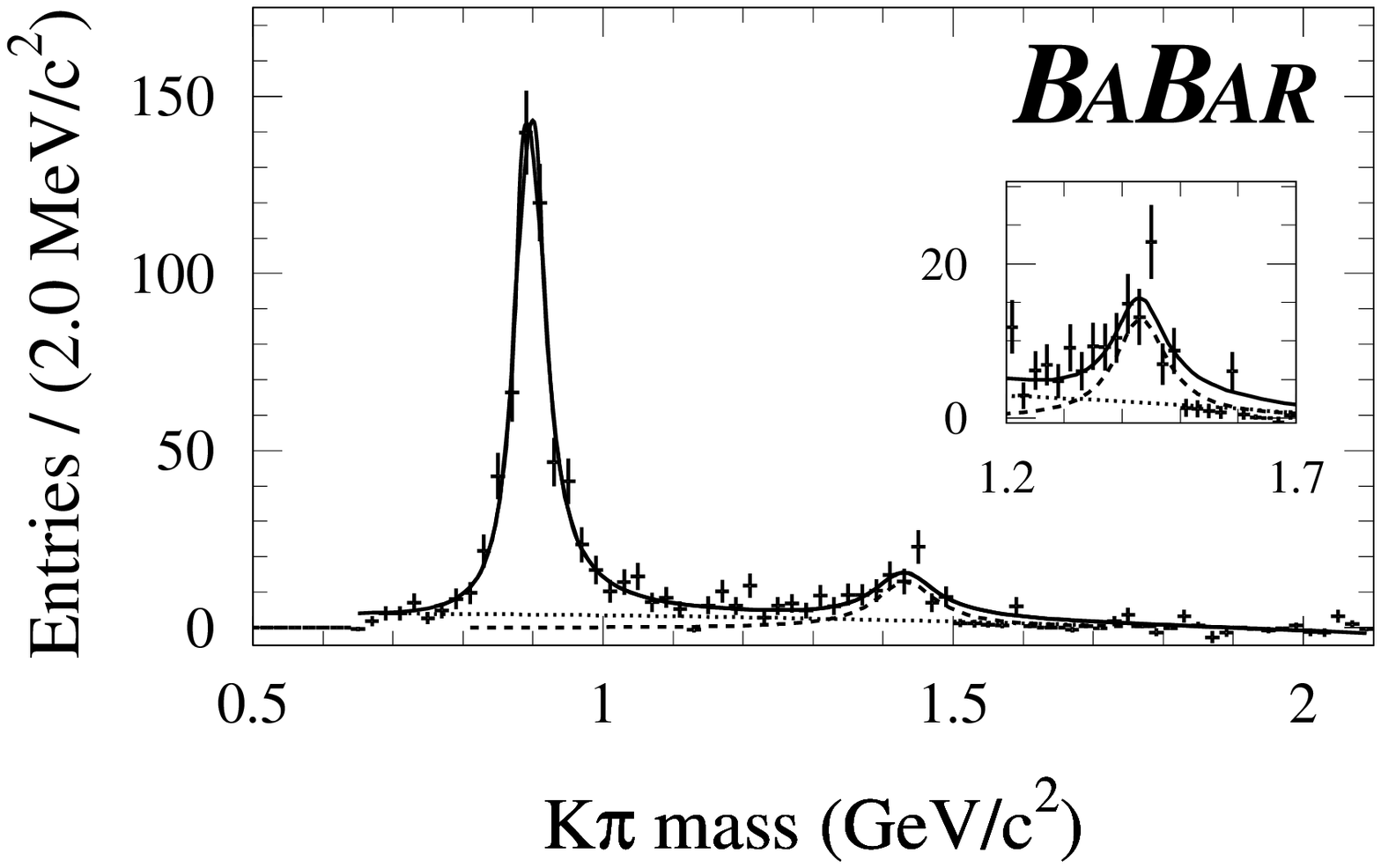,width=4.5in}
\caption{Background-subtracted $K\pi$ mass distribution in 
	the mode $\Btopsikstz$, $\ksttokpi$. }
\label{fig:heavykstar}
\end{figure}

\subsection{Angular Distribution and Fitting Method}

The transversity frame is defined in the $\jpsi$ rest frame.
The $\Kstar$ direction defines the $-x$ axis.
The $K\pi$ plane defines the $(x,y)$ plane, with $y$ such 
that $p_y(K) > 0$. The $z$ axis is
normal to that plane so that the $(x,y,z)$ frame is 
right handed. The transversity angles $\thetatr$ and $\phitr$ 
are defined as the polar and azimuthal angles of the $\ell^+$ 
originating from the $\jpsi$. The $\Kstar$ helicity angle 
$\thetakstar$ is the angle between the $K$ direction and the 
direction opposite to the $\jpsi$ in the $\Kstar$ rest frame.
Defining:
\begin{eqnarray*}
f_1 & = & + \ \, 9/(32\pi)\cdot\,  
		2\cq{\thetakstar}(1-\sq{\thetatr}\cq{\phitr}) \\
f_2 & = & + \ \, 9/(32\pi)\cdot\, 
		\sq{\thetakstar}(1-\sq{\thetatr}\sq{\phitr}) \\
f_3 & = & + \ \, 9/(32\pi)\cdot\, 
		\sq{\thetakstar}\sq{\thetatr} \\
f_4 & = & + \ \, 9/(32\pi)\cdot\, 
		\sq{\thetakstar}\sin{2\thetatr}\sphitr\cdot\zeta \\
f_5 & = & - \ \, 9/(32\pi)\cdot\,  
		1/\sqrt{2}\cdot\sin{2\thetakstar}
			\sq{\thetatr}\sin{2\phitr} \\
f_6 & = & + \ \, 9/(32\pi)\cdot\,  
		1/\sqrt{2}\cdot\sin{2\thetakstar}
			\sin{2\thetatr}\cos{\phitr}\cdot\zeta~,
\end{eqnarray*}
the angular distribution $g(\cthetatr, \cthetakstar, \phitr)$ reads
\begin{eqnarray}
\label{eqn:distrib}
g & \,=\, & \gfrac{1}{\Gamma}\gfrac{\dd^{3}
		\Gamma}{\dd\cthetatr\;\,\dd\cthetakstar\;\,\dd\phitr} \\
	& \,=\, & f_1\azd  + f_2\apd  + f_3\atd   
	 +  f_4\pipt + f_5\przp + f_6\pizt~.   \nonumber
\end{eqnarray}
The parameter $\zeta$ is $+1$ for $\Bu$ and $\Bz$ and $-1$ for 
$\Bub$ and $\Bzb$. For the $\CP$ eigenstate $\KS\piz$, a 
$\Bz\Bzb$ from the decay of an $\FourS$ has $\zeta=\pm 1/(1+x_d^2)$.
However, the initial $\Bz$ flavor is not determined in this analysis so
the appropriate value for $\zeta$ is the average, namely zero.
\vs
The amplitudes are fitted by means of a UML fit
exploiting the angular and $\mes$ information.
All $\B$ candidates within $5.2<\mes<5.3~\gevcc$
enter the fit, where the $\mes$ sidebands are used to normalize
non $\Btopsikst$ background. The biases on the amplitudes due 
to limited acceptance and background are corrected considering only 
the projections on the basis of the $f_i$'s~\cite{ref:samir}.
\vs
Signal events are described by the PDF
\begin{equation}
	\gobs = g(\vec{\omega}_j)
	\times\epsilon(\vec{\omega}_j)/\langle\epsilon\rangle~,
\end{equation}
where $\vec{\omega_j}$ stands for the three angles of the 
observed event $j$, and $\epsilon(\vec{\omega}_j)$
is the efficiency for this event. Rewriting 
Eq.~(\ref{eqn:distrib}) as $g = \sum_{i=1}^{6} f_i A_i^2$, 
where the $A_i^2 (i=1,6)$ are $\azd$, $\apd$, $\atd$, $\pipt$, 
$\przp$ and $\pizt$, the mean efficiency 
$\langle\epsilon\rangle$ reads
\begin{equation}
	\langle\epsilon\rangle = \int g\times \epsilon 
		= \sum_{i=1}^{6}A_i^2\xi_i~,
\end{equation}
where the $\xi_i = \int f_i\times \epsilon$ 
are amplitude independent.
The signal part of the log-likelihood
\begin{equation}
{\cal L}_{\rm signal} = \sum_{j=1}^{N_{\rm obs}} 
		\log\left(\gobs(\vec{\omega}_j)\right)~,
\end{equation}
where $N_{\rm obs}$ is the number of observed events, becomes
\begin{equation}
{\cal L}_{\rm signal} = \sum_{j=1}^{N_{\rm obs}}
		\log\left(g(\vec{\omega}_j)\right) 
+\sum_{j=1}^{N_{\rm obs}}\log\left(\epsilon(\vec{\omega}_j)\right)
	            -N_{\rm obs} 
			\log\left(\sum_{i=1}^{6}A_i^2\xi_i\right)~.
\end{equation}
Since the $\epsilon(\vec{\omega}_j)$ do not depend on the 
amplitudes, the term 
$\sum_{j=1}^{N_{\rm obs}}\log(\epsilon(\vec{\omega}_j))$ 
is left out: detailed knowledge of the acceptance is 
not necessary in this approach.
\vs
The coefficients $\xi_i$ are evaluated using a detailed
simulation of the \babar\ detector. Separate series 
of $\xi_i$'s are used for each $\Btopsikst$ channel. The 
values found for $\xi_i$, $(i=1,3)$, are close to the efficiency 
of the final state, with a lower value for $\xi_1$, especially
in modes involving $\piz$, due to the requirement on 
$\cthetakstar$. The $\xi_i$, $(i=4,6)$, are compatible with zero.
\vs
The PDF of the non $\Btopsikst$ background 
$g_B^{\rm obs}(\vec{\omega}_j)$ is written as a distribution $g_B$
of amplitudes $B^2_i (i=1,6)$: 
$g_B^{\rm obs}(\vec{\omega}_j) = g_B(\vec{\omega}_j)
\times\epsilon(\vec{\omega}_j)/\langle\epsilon\rangle$.
The shape and level of the (S)CF background is amplitude 
dependent. The bias due to (S)CF background can be
corrected by a modification $\tilde{\xi}_i$ of the $\xi_i$,
obtained by integrating, in addition to the signal as before,
over the (S)CF contributions. In contrast to $\xi_i$, 
the $\tilde{\xi}_i$ depend on the amplitudes used
in the simulation. The maximum bias on the fitted 
amplitudes was found to be a few parts per thousand.
\vs
Finally, the complete log-likelihood has the form
\begin{eqnarray}
{\cal L} &= & \sum_{j=1}^{N_{\rm obs}}
	\log
	\bigg( x\cdot G(\mes_j) g(\vec{\omega}_j) 
	 \;+\;(1-x)\cdot B(\mes_j)  g_B(\vec{\omega}_j)\bigg) 
		\nonumber\\
	 &&\hspace{0.1cm}
             -\; N_{\rm obs} \log 
		\left(\sum_{i=1}^{6}   
			\tilde{\xi}_i\left( x\cdot  A_i^2 +
			(1-x)\cdot B_i^2 \right) \right) - {\cal N}~,
	\nonumber
\end{eqnarray}
where $G(\mes)$ and $B(\mes)$ are the Gaussian signal function
and ARGUS background function, respectively.
The parameter $x$ measures the signal fraction.
The normalization of the amplitudes is enforced by the 
likelihood extension ${\cal N} =N_{\rm obs}(\azd+\apd+\atd)$.
\begin{table}[t]
\caption{Systematic uncertainties. Details are given in the text.}
\begin{center}
\begin{tabular}{lccccc}  \hline
					&	$\azd$	&	 $\atd$	&	$\apd$	&	$\phit$	&	$\phip$	\\ \hline
	MC stat.		&       0.006   &       0.006	&       0.007	&       0.04	&       0.06	\\
	Backgr.			&	0.002	&	0.005	&	0.006	&	0.01	&	0.00	\\
	Track. \& PID		&	0.002	&	0.006	&	0.004	&	0.00	&	0.02	\\
	Heavy $\Kstar$			&	0.005	&	0.035	&	0.031	&	0.04	&	0.02	\\ \hline
	Total				&	0.008	&	0.036	&	0.033	&	0.06	&	0.07 \\\hline
\label{tab:systematics}
\end{tabular}
\end{center}
\vspace{0.5cm}
\setlength{\tabcolsep}{1.3pc}
\vspace*{-0.6cm}
\caption{Preliminary amplitude moduli-squared and phases. The
	uncertainties are statistical and systematic.}
\begin{center}
\begin{tabular}{lc}  \hline
	Quantity &	\hspace{0.0cm}		
	 			Value			\\ \hline
	$\azd$	 &				
	 	0.597	 $\pm$ 0.028 	$\pm$ 0.008	\\	
	$\atd$	 &				
		0.160	 $\pm$ 0.032 	$\pm$ 0.036	\\
	$\apd$	 &				
		0.243	 $\pm$ 0.034 	$\pm$ 0.033	\\ \hline
	$\phit=\arg(\at/\az)$	 &		
		$-0.17$	 $\pm$ 0.16  	$\pm$ 0.06	\\
	$\phip=\arg(\ap/\az)$	 &		
		2.50 	 $\pm$ 0.20  	$\pm$ 0.07	\\ \hline
\label{tab:results}
\end{tabular}
\end{center}
\end{table}
\begin{figure}
\center
\epsfig{file=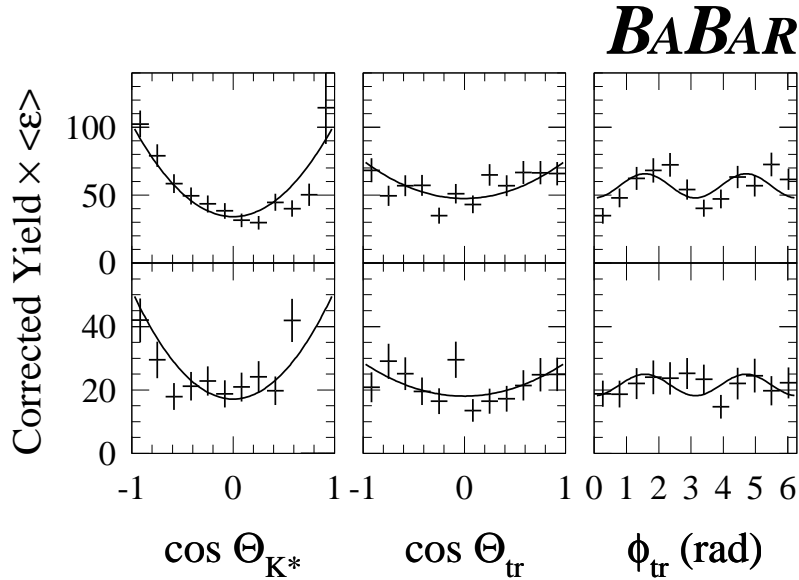,width=4.5in}
\caption{One-dimensional projections of the three transversity 
	angles. The result of the fit is superimposed.
	The data are corrected for bin-by-bin acceptance. 
	Top plot: modes without $\piz$ in the final state, and
	bottom plot: modes with $\piz$.}
\label{fig:fitdata}
\end{figure}

\subsection{Systematic Uncertainties}

The limited MC statistics (32000 events per channel) gives 
rise to a systematic uncertainty on acceptance and cross 
feed corrections. A fraction of the feed-across 
background may be absorbed in the fit by the non $\Btopsikst$ 
background component. The size of this effect 
is evaluated on MC samples, where a set of fits are performed 
with the non-$\jpsi K^*$ background (ARGUS) function removed.  
The largest deviation found is taken as an estimate of the 
associated systematic uncertainty. Discrepancies between 
the true detector and the simulation affect the acceptance 
correction. The scale of the effect is estimated by varying
the tracking and particle identification efficiencies at 
MC level. Non-resonant $K\pi$ combinations, higher mass $K\pi$ 
resonances and their interference with the $\Kstar(892)$ 
can contribute to the events in the $\Kstar$ mass window.  
The size of the potential bias from these events is estimated
by the magnitude of the shift in the fitted angular amplitudes 
when events in the region $|m_{K\pi}-1.2|<0.1 \gevcc$ are 
included in the fit. 
\vs
Table~\ref{tab:systematics} quantifies the systematic effects
of the contributions listed above on the fit observables.

\subsection{Results}

The UML fit results for amplitude moduli-squared and phases 
are given in Tab.~\ref{tab:results}.
The errors account for statistical and systematic uncertainties.
One-dimensional projections of the fitted amplitude
spectrum on the three transversity angles are shown
in Fig.~\ref{fig:fitdata} for modes without $\piz$ (upper
plots) and with $\piz$ (lower plots) in the final states.
Figure~\ref{fig:triangle} visualizes the quantitative effects 
of the different corrections on the amplitude moduli-squared. 
Only the acceptance correction exceeds the range defined by 
the statistical uncertainties.
\begin{figure}[p]
\center
\epsfig{file=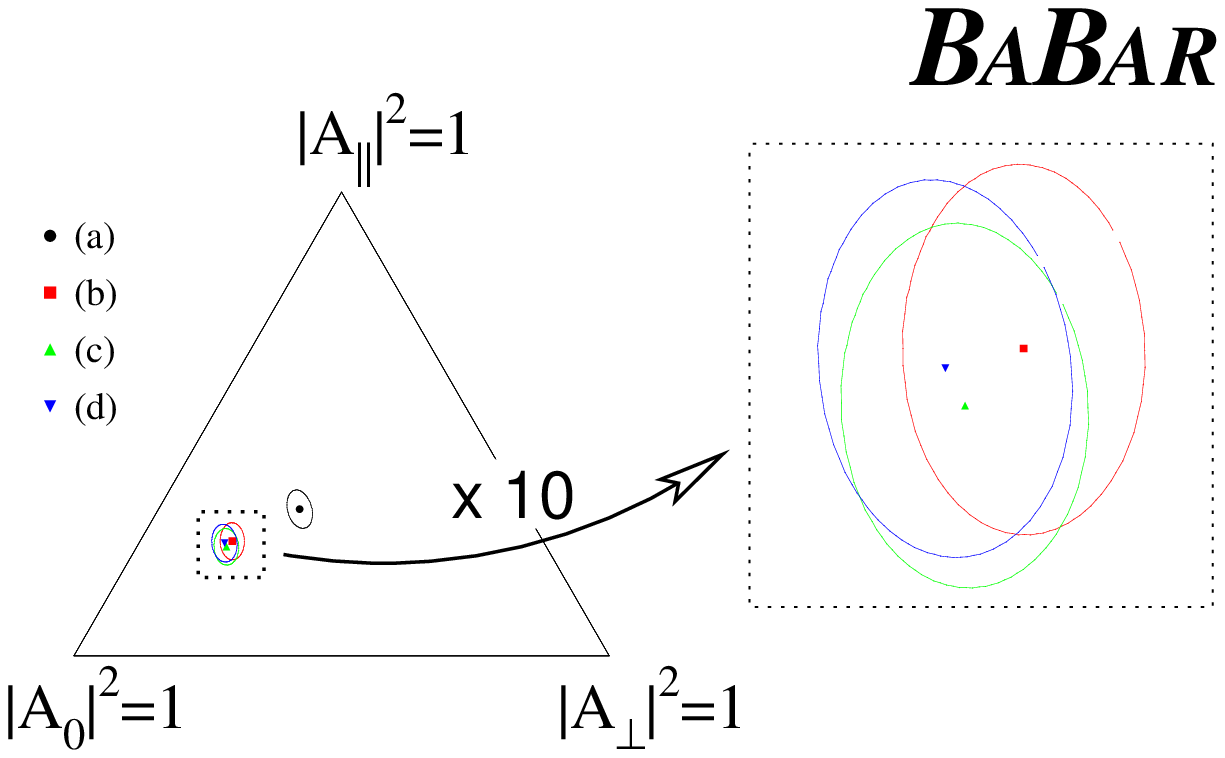,width=4.0in}
\caption{Results of the amplitude fits and effects 
	of the corrections applied. The results are shown 
	in the allowed domain of the amplitude moduli-squared 
	defined by the conditions $\azd + \apd + \atd = 1$ 
	and $|A_i|^2 \ge 0, (i=0, \parallel, \perp)$. Left: 
	the entire allowed domain, right: zoom by a factor of ten
	into the (b, c, d) region. (a), angular fit with no 
	acceptance correction or $\mes$ dependence, 
	(b), acceptance correction applied, (c), $\mes$ 
	dependence added, and (d), (S)CF correction added.}
\label{fig:triangle}
\vspace{2cm}
\center
\epsfig{file=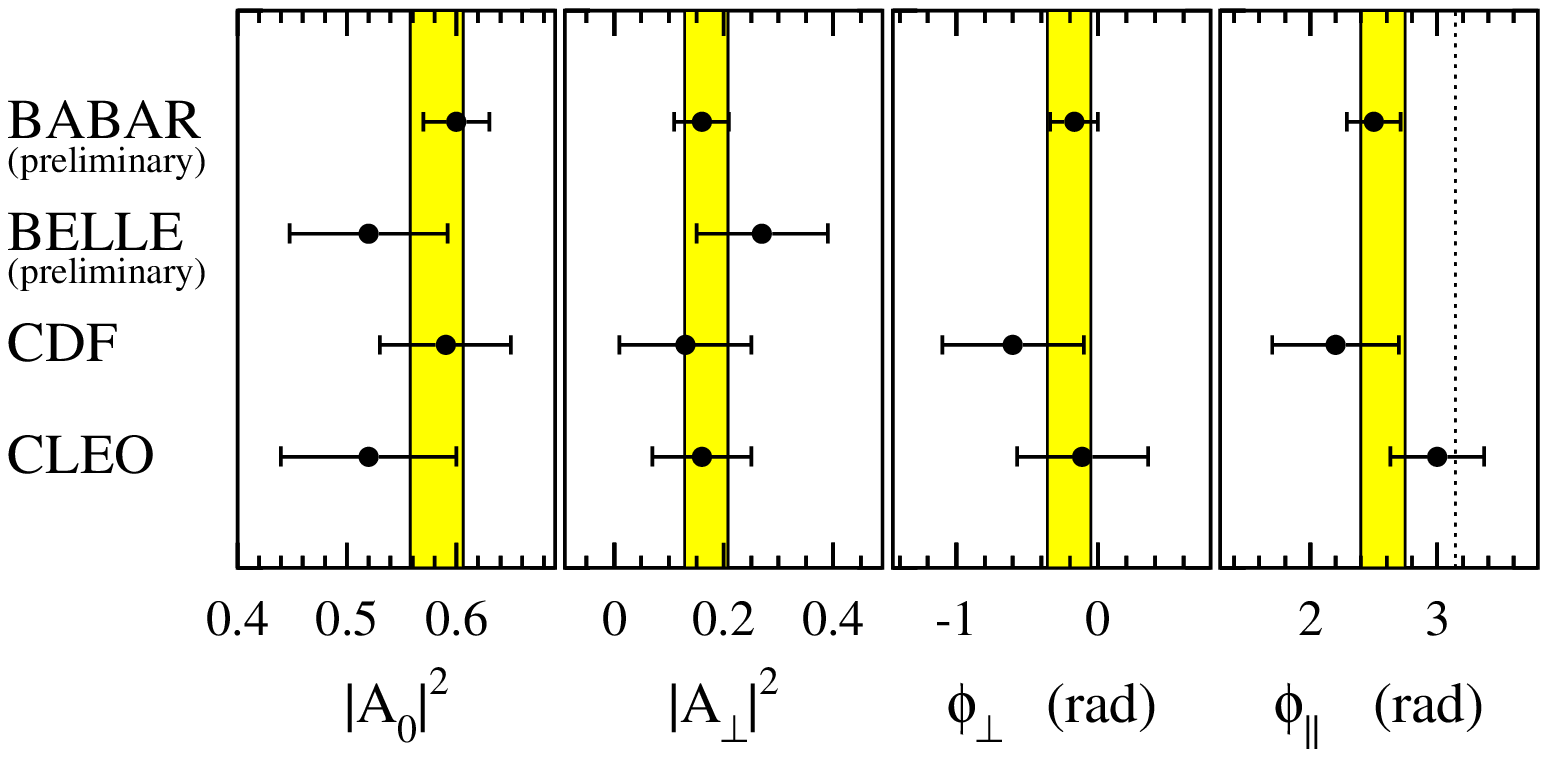,width=4.5in}
\caption{Comparison of measured 
	amplitude moduli-squared and phases (see text for
	references).
	The Belle fit assumed no final-state interaction.
	The shaded bands give the averages.}
\label{fig:jkcomp}
\end{figure}
Figure~\ref{fig:jkcomp} shows the comparison between 
this preliminary measurement and results from 
BELLE~\cite{ref:jkbelle} (preliminary), CDF~\cite{ref:jkcdf} and 
CLEO~\cite{ref:jkcleo}, which are found to be consistent.
The shaded bands indicate the weighted averages.
The $\babar$ values improve the precision by about a factor 
of two. 
\vs
We find a dominant longitudinal component and a 
small $P$ wave contribution. If $\stwob$ is measured 
from a time dependent $\CP$ asymmetry 
analysis in the mode $\B\to\jpsi\KS\piz$, the dilution 
introduced amounts to $D_\perp = (1-2\atd) = 0.68\pm 0.10$. 
The uncertainty on $\atd$ contributes with relative 14\% 
to the total error of $\stwob$. This measurement of 
$\phip$ implies a deviation from $\pi$ by
3 standard deviations (3.4 for the world average), and 
thus represents evidence for non-vanishing final-state 
interaction.

\subsection*{Acknowledgements}

I am indebted to all my $\babar$ colleagues who helped me
preparing this talk. Many thanks to the experts
D.~Bernard, G.~Cavoto,
C.~Dallapiccola, J.~Olsen and M.~Verderi for their knowledgeable
comments on the analyses reported here.
We are grateful for the contributions of our \pep2\ colleagues in
achieving the excellent luminosity and machine conditions
that have made this work possible.
We acknowledge support from the
Natural Sciences and Engineering Research Council (Canada),
Institute of High Energy Physics (China),
Commissariat \`a l'Energie Atomique and
Institut National de Physique Nucl\'eaire et de Physique des Particules
(France),
Bundesministerium f\"ur Bildung und Forschung
(Germany),
Istituto Nazionale di Fisica Nucleare (Italy),
The Research Council of Norway,
Ministry of Science and Technology of the Russian Federation,
Particle Physics and Astronomy Research Council (United Kingdom), the
Department of Energy (US),
and the National Science Foundation (US). In addition, individual support 
has been received from the Swiss 
National Foundation, the A. P. Sloan Foundation, the Research Corporation,
and the Alexander von Humboldt Foundation.
The visiting groups wish to thank 
SLAC for the support and kind hospitality
extended to them.


\end{document}